\documentstyle[12pt,epsfig]{article}

\input epsf

\newcommand{\ppp}[1]{%
        \setbox0=\hbox{#1}%
        \kern-.02em\copy0\kern-\wd0
        \kern+.04em\copy0\kern-\wd0
        \kern-.02em\raise.0217em\box0}

\newcommand{\lsim}{
 \mathrel{\setbox0=\hbox{$<$}\raise0.6ex\copy0\kern-\wd0
 \lower0.65ex\hbox{$\sim$}}}

\newcommand{\gsim}{
 \mathrel{\setbox0=\hbox{$>$}\raise0.6ex\copy0\kern-\wd0
 \lower0.65ex\hbox{$\sim$}}}

\newcommand{\PRD}[3]{Phys.\ Rev.\ D {\bf {#1}}, {#2} ({#3})}

\newcommand{\PRL}[3]{Phys.\ Rev.\ Lett.\ {\bf {#1}}, {#2} ({#3})}

\newcommand{\NPB}[3]{Nucl.\ Phys.\ B {\bf {#1}}, {#2} ({#3})}
\newcommand{\PLB}[3]{Phys.\ Lett.\ B {\bf {#1}}, {#2} ({#3})}
\newcommand{\ZPC}[3]{Z. Phys.\ C {\bf {#1}}, {#2} ({#3})}
\newcommand{\JPG}[3]{J. Phys.\ G {\bf {#1}}, {#2} ({#3})}
\newcommand{\EJP}[3]{Eur. J. Phys.\ C {\bf {#1}}, {#2} ({#3})}

\begin{document}
%

\begin{titlepage}
\renewcommand{\thefootnote}{\fnsymbol{footnote}}
\makebox[2cm]{}\\[-1in]
\begin{flushright}
\begin{tabular}{l}
TUM/T39-99-6
\end{tabular}
\end{flushright}
\vskip0.4cm
\begin{center}
  {\Large\bf Comments on Exclusive Electroproduction of 
\\[0.2cm]
Transversely Polarized Vector Mesons\footnote{Work 
supported in part by BMBF}
}\\ 

\vspace{2cm}
L. Mankiewicz$^{a b}$ and G. Piller$^a$\\

\vspace{1.5cm}

\begin{center}

{\em$^a$Physik Department, Technische Universit\"{a}t M\"{u}nchen, 
D-85747 Garching, Germany}

{\em $^b$N. Copernicus Astronomical Center, ul. Bartycka 18,
PL--00-716 Warsaw, Poland}

\end{center}

\vspace{1cm}


\vspace{3cm}

\centerline{\bf Abstract}
\begin{center}
\begin{minipage}{15cm}

We discuss the electroproduction of light vector mesons 
from transversely polarized photons.  
Here QCD factorization cannot be applied as shown 
explicitly in a leading order calculation of corresponding Feynman 
diagrams. 
It is emphasized that present infrared  singular contributions  
cannot be regularized through phenomenological meson distribution 
amplitudes with suppressed endpoint configurations. 
We point out that infrared divergencies arise also from integrals 
over skewed parton distributions of the nucleons.

In a phenomenological analysis of transverse 
vector  meson production  model dependent regularizations  
have to be applied. If this procedure preserves 
the analytic structure suggested by a leading order calculation of  
Feynman diagrams, one obtains contributions from nucleon 
parton distributions and their derivatives.
In particular polarized gluons enter  only through their derivative.

\end{minipage}
\end{center}

\end{center}
\end{titlepage}
\setcounter{footnote}{0}

\newpage

In recent years exclusive electroproduction of mesons from nucleons  
has become a topic of broad interest. 
Experimental and theoretical advances have supported 
this development. 
At high energies a large amount of data has become available 
from experiments at CERN (NMC) 
and DESY (HERA, HERMES) 
(for references see e.g. \cite{HERA,HERMES3}).  
Further measurements are carried out at DESY, and 
discussed at CERN (COMPASS) \cite{COMPASS} and 
TJNAF \cite{TJNAF}. From the theoretical side  
a factorization theorem proven in 
ref.\cite{CFS97} let the basis for many investigations. 
It states that the underlying photon-parton sub-processes 
are dominated for longitudinally polarized photons and large 
photon virtualities, $Q^2 \gg \Lambda_{\rm{QCD}}^2$, 
by short distances and, hence, can be calculated perturbatively. 

As already emphasized in \cite{CFS97}, the interaction of 
transversely polarized photons cannot be treated in a 
framework based on QCD factorization. 
This is due to infrared sensitive contributions which 
e.g. result from large size quark-antiquark 
configurations in the produced meson. 
A straightforward QCD analysis of 
transverse vector meson production, as done  
for longitudinal ones (see  
refs.\cite{Rys93,FS94,Rad97,Hood98,MPW98,Mankiewicz:1999aa,Vanderh98}), 
is therefore not possible. 
On the other hand, expected new data will provide more detailed information 
on these processes and help to investigate 
strong interaction dynamics at large distances.

Descriptions of exclusive meson production 
from transversely polarized photons rely on model assumptions which are 
needed to regularize the present, at least  logarithmic, 
infrared singularities.
In this note we outline several important properties of the involved  
production amplitudes:
(i) different vector meson distribution amplitudes which 
enter in the production amplitude are related by 
Wandura-Wilczek type relations based on Lorentz invariance
\cite{BallBraun96,BallBraun98}. 
These have significant implications for the nature of the 
existing infrared singularities.
(ii) Infrared divergencies arise also from integrals 
over skewed parton distributions of nucleons.
(ii) A model calculation of the vector meson production 
amplitude guided by Feynman diagrams yields an 
analytic structure which is richer than for meson production 
from longitudinal photons. 
In particular one  obtains  
contributions which involve both parton distributions and 
their derivatives. 

In the following we restrict ourselves to 
$\rho$ production. An important ingredience of the corresponding 
amplitude are the  meson distribution amplitudes. 
Following refs.\cite{BallBraun96,BallBraun98} we parametrize the  
meson-to-vacuum
matrix elements of vector and axial vector 
\footnote{We use the convention 
of \cite{ItzZub} for $\gamma_5$ and the epsilon tensor.} 
currents as:  
\begin{eqnarray}\label{eq:vector_mel}
\langle 0 | {\bar q}(0)[0;x]\gamma_\mu q(x) |\rho(p,\lambda)\rangle =
&p_\mu& \frac{e^{(\lambda)} \cdot x}{p \cdot x} f_\rho m_\rho
\int_0^1 d\tau \, e^{-i\tau p\cdot x} \phi_{||}(\tau) 
\nonumber \\
&+& \left( e^{(\lambda)}_\mu - p_\mu \frac{e^{(\lambda)} \cdot x}{p \cdot x} 
\right) f_\rho m_\rho
\int_0^1 d\tau \, e^{-i\tau p\cdot x} g^v(\tau) \, ,
\end{eqnarray}
and
\begin{equation}\label{eq:a_vector_mel}
\langle 0 | {\bar q}(0)[0;x]\gamma_\mu \gamma_5 q(x) |\rho(p,\lambda)\rangle =
\frac{1}{4} \varepsilon_{\mu\nu\rho\sigma} e^{(\lambda)\, \nu} p^\rho
x^\sigma 
f_\rho m_\rho \int_0^1 d\tau \, e^{-i\tau p\cdot x} g^a(\tau) \, .
\end{equation}
Here $p^{\mu}$ is the four-momentum of the $\rho$ meson with 
invariant mass $m_\rho$ and decay constant $f_\rho$.  
The vector meson polarization is specified by $\lambda$ 
which corresponds to the polarization vector $e^{(\lambda)}$. 
The light-cone matrix elements, as well as the distribution amplitudes 
in eqs.(\ref{eq:vector_mel},\ref{eq:a_vector_mel}), are defined at 
a certain renormalization scale $\mu$ which we suppress if convenient. 
Gauge invariance is guaranteed by the path-ordered exponential  
$$
[0;x] = \,
{\cal P} \exp [ -i g x_{\mu} 
\-\int_0^1  A^{\mu}(x \eta) \, d \eta]
$$  
which reduces to 
$1$
in  axial gauge $n\cdot A=0$ 
($g$ stands for the strong coupling constant and 
$A^{\mu}$ denotes the gluon field). In light-cone gauge the twist-$2$
distribution amplitude $\phi_{||}(\tau)$  can be related to the wave
function of the minimal quark-antiquark Fock state in a longitudinally
polarized  meson \cite{ER78,BL79,CZ84}. The twist-$2$ distribution
amplitude for a transversally polarized $\rho$ 
is determined by the matrix
element of a chiral-odd tensor quark operator and does not
contribute to the process considered here. 
In the following we  use also the antisymmetric 
distribution $\Phi_{||}(\tau)$ given by \cite{BallBraun96}: 
\begin{equation}
\Phi_{||}(\tau) =
\frac{1}{2}\left[ {\bar \tau}\int_0^\tau du \, \frac{\phi_{||}(u)}{{\bar u}}  
- \tau \int_\tau^1 du \, \frac{\phi_{||}(u)}{u}\right] \, ,
\end{equation}
with $\bar \tau = 1-\tau$. 

As explained in \cite{BallBraun96,BallBraun98} 
twist-$2$ and twist-$3$ string operators contribute to the distributions
$g^v(\tau)$ and $g^a(\tau)$. Lorentz invariance leads to the presence 
of twist-$2$
contributions which are given by Wandzura-Wilczek type relations:
\begin{eqnarray}\label{eq:Wandzura_Wilczek1}
g^v(\tau) &=& 
\frac{1}{2}\left[ \int_0^\tau du \, \frac{\phi_{||}(u)}{{\bar u}}  
+ \int_\tau^1 du \, \frac{\phi_{||}(u)}{u}\right] \, ,
\\
\label{eq:Wandzura_Wilczek2}
g^a(\tau) &=& 
2 \left[ {\bar \tau} \int_0^\tau du \, \frac{\phi_{||}(u)}{{\bar u}}  
+ \tau \int_\tau^1 du \, \frac{\phi_{||}(u)}{u}\right] \, .
\end{eqnarray}
Both, $g^v(\tau)$ and $g^a(\tau)$ are symmetric functions of $\tau$.
Twist-$3$
contributions are related to matrix elements of three-particle
quark-gluon-quark operators \cite{BallBraun98} and will not be considered
here. 

In general one can assume  
$\sigma_{\phi}=\int_0^1 du \, {\phi_{||}(u)}/{u}$ is different from zero. 
This is quite natural since $\sigma_{\phi}=0$ can  be 
fulfilled, if at all,  only at one particular scale $\mu$ 
due to the scale dependence of $\phi_{||}(u;\mu)$ \cite{BallBraun96}. 
As a consequence one finds from 
eqs.(\ref{eq:Wandzura_Wilczek1},\ref{eq:Wandzura_Wilczek2}) 
in the limit $\tau \rightarrow 0$: 
$g^v(\tau) \sim \sigma_{\phi}$,  
$g^a(\tau) \sim \sigma_{\phi} \, \tau$ and 
$\Phi_{||}(\tau) \sim \sigma_{\phi} \, \tau$.

The amplitude ${\cal M}^{\gamma^*_\perp \to \rho_\perp}$ for $\rho$ meson 
production from transversally polarized virtual photon can be split into 
parts,
\begin{equation}\label{eq:production_amplitude}
{\cal M}^{\gamma^*_\perp \to \rho_\perp} =
{\cal M}^{\gamma^*_\perp \to \rho_\perp}_G +
{\cal M}^{\gamma^*_\perp \to \rho_\perp}_q +
{\cal M}^{\gamma^*_\perp \to \rho_\perp}_{\Delta G} +
{\cal M}^{\gamma^*_\perp \to \rho_\perp}_{\Delta q}, 
\end{equation}
involving unpolarized and polarized generalized quark 
and gluon distribution functions, respectively. 
For convenience we concentrate in the following on the 
gluon contributions.
A straightforward calculation of leading order Feynman diagrams 
along the lines of ref.\cite{MPW98} gives
for the amplitude which involves the unpolarized gluon 
distribution:
\begin{eqnarray}\label{eq:amplitude_gl_unpol}
{\cal M}^{\gamma^*_\perp \to \rho_\perp}_G &=&
i \frac{g^2}{32 N_C} \,\frac{f_\rho m_\rho}{{\bar Q}^2}
\frac{\bar{N}(P',S') \hat{n} N(P,S)}{{\bar P} \cdot n}  
\,\,E \cdot e^*
\Bigg{\{}
\nonumber\\
&&
{\cal I}_1\,\int_{-1}^{1}du\, G(u,\xi) 
\left[ \frac{\bar \omega}{(\xi-u-i \epsilon)}+
\frac{\bar \omega}{(\xi+u-i \epsilon)} \right] 
\nonumber \\
&+&
{\cal I}_2\,\int_{-1}^{1}du\, G(u,\xi) 
\left[ \frac{1}{(\xi-u-i \epsilon)^2}+
\frac{1}{(\xi+u-i \epsilon)^2} \right] \Bigg{\}}\,,
\end{eqnarray}
while the contribution of the polarized generalized gluon distribution
reads:
\begin{eqnarray}\label{eq:amplitude_deltaG}
{\cal M}^{\gamma^*_\perp \to \rho_\perp}_{\Delta G} &=&
-\frac{g^2}{32 N_C} \, \frac{f_\rho m_\rho}{{\bar Q}^2}
\frac{\bar{N}(P',S') \gamma_5 \hat{n} N(P,S)}{{\bar P} \cdot n}
\,\, 
\varepsilon_{\mu\nu\alpha\beta} E^\mu e^{\nu\,*}
\frac{n^\alpha n^{*\beta}}{n \cdot n^*}
\nonumber \\
&\times& 
{\cal I}_2 \,\,
\int_{-1}^{1}du\, \Delta G(u,\xi) 
\left[ \frac{1}{(\xi-u-i \epsilon)^2}-
\frac{1}{(\xi+u-i \epsilon)^2} \right] \,.
\end{eqnarray}
Here $E^\mu$ denotes the polarization vector of the virtual
photon. $N(P,S)$ and $\bar N(P',S')$  are the  Dirac spinors   
of the initial and scattered nucleon, respectively, 
with the corresponding four-momenta  $P$, $P'$  
and spins $S$, $S'$. 
The average nucleon momentum is  $\bar P = (P + P')/2$, 
and the momentum transfer is  $r = P-P'$. 
The produced  meson carries the four-momentum 
$q'$ and $\bar q = (q + q')/2$. 
Furthermore, we have introduced the variables  $\bar Q^2 = - \bar q^2$, 
$\bar \omega = 2 \bar q \cdot \bar P/ (- \bar q^2)$ and 
$\xi = 1/\bar \omega$. 
Finally, $n$ is  a light-like vector with
$n\cdot a = a^+ = a^0+ a^3$ for any vector $a$, and
$\hat n = \gamma_{\mu} n^{\mu}$. 
The $\rho$ meson distribution amplitudes 
(\ref{eq:vector_mel},\ref{eq:a_vector_mel}) 
determine the integrals ${\cal I}_{1/2}$ as 
given explicitly in eq.(\ref{eq:I1_I2}). 
$G(u,\xi)$ and $\Delta G(u,\xi)$ stand for  the skewed unpolarized
and polarized gluon distributions \cite{Ji97,Rad97}. 
In the forward limit, $\xi \to 0$, they 
reduce to the ordinary unpolarized  and polarized 
gluon distributions of the nucleon: 
\begin{eqnarray}\label{eq:forward_limit}
&&\lim_{\xi \to 0} G(u,\xi) = u g(u) \, ,
\nonumber \\
&&\lim_{\xi \to 0} \Delta G(u,\xi) = u \Delta g(u) \, .
\end{eqnarray}

For simplicity we have omitted in 
eqs.(\ref{eq:amplitude_gl_unpol},\ref{eq:amplitude_deltaG})
so-called $K$-terms \cite{Ji97,Rad97}. 
Their contributions to the production amplitudes 
(\ref{eq:amplitude_gl_unpol},\ref{eq:amplitude_deltaG}) can be obtained simply by replacing
the skewed gluon distributions  by 
$K$-distributions (including corresponding Dirac pre-factors) 
\cite{MPW98}. 

The integrals ${\cal I}_1$ and ${\cal I}_2$ contain  the dependence of the 
production amplitudes on the meson distributions $g^a$, $g^v$ and 
$\Phi_{||}$: 
\begin{eqnarray}\label{eq:I1_I2}
{\cal I}_1 &=& \int_0^1 \frac{d\tau}{\tau}\, \left( 4 g^v(\tau) - 2
\frac{\Phi_{||}(\tau)}{\tau} + 
\frac{g^a(\tau)}{2 \tau {\bar \tau}} \right),
\nonumber \\
{\cal I}_2 &=& \int_0^1 \frac{d\tau}{\tau}\, \left( 2 g^v(\tau) + 
\frac{g^a(\tau)}{2 \tau {\bar \tau}} \right) \, .
\end{eqnarray}
Both, ${\cal I}_1$ and ${\cal I}_2$ are divergent due to the behavior
of the integrands at $\tau \to 0$. These infrared
divergences make  
QCD factorization impossible \cite{CFS97,Rad97}. 
Currently no QCD framework is available to 
deal with this problem of infrared sensitive contributions.
As a consequence all predictions for $\rho$  production  
from transversely polarized photons are  model-dependent. 
One of the points we want to emphasize in this note is, 
that a modification of the 
end-point behavior of the the twist-$2$ amplitude
$\phi_{||}(\tau)$ cannot  cure the infrared 
divergence discussed above. 
This is a direct consequence of the Ball-Braun relation 
(\ref{eq:Wandzura_Wilczek1},\ref{eq:Wandzura_Wilczek2}). 

For the asymptotic distribution amplitude  
$\phi_{||}(\tau) = 6 \tau {\bar \tau}$  the integrands  in 
eq.(\ref{eq:I1_I2}) are proportional to $12/\tau$ and $9/\tau$, respectively. 
One, therefore, might expect 
that any phenomenological regularization which is applied to render the
integrals finite, see eg. \cite{MarRysTeub97,Rys97}, 
leads to a ratio 
${\cal I}_1/{\cal I}_2$ close to one.

Another important point is the  dependence of the 
production amplitudes on the unpolarized
and polarized gluon distributions. 
According to eq.(\ref{eq:amplitude_gl_unpol}) the unpolarized skewed gluon
distribution enters ${\cal M}^{\gamma^*_\perp \to \rho_\perp}_G$
through
\begin{equation} \label{eq:G_old}
\int_{-1}^{1}du\, G(u,\xi) 
\left[ \frac{1}{(\xi-u-i \epsilon)}+
\frac{1}{(\xi+u-i \epsilon)} \right]\,,
\end{equation}
\noindent and
\begin{equation} \label{eq:G_new}
\int_{-1}^{1}du\, G(u,\xi) 
\left[ \frac{1}{(\xi-u-i \epsilon)^2}+
\frac{1}{(\xi+u-i \epsilon)^2} \right] \, .
\end{equation}
The first integral is also present in the leading-twist
production amplitude of $\rho$ mesons via 
longitudinally polarized  photons 
(see e.g. \cite{MPW98}). 
The second contribution, which involves the square 
of $(\xi \pm u-i \epsilon)$ in the denominator, has not been 
considered before \cite{MarRysTeub97}.
We believe that any model of transverse $\rho$ production should include  
both contributions. 
Note that the integrals (\ref{eq:G_old}) and (\ref{eq:G_new}) are well defined 
only if $G(u,\xi)$ and its first derivative are 
continuous at $u = \xi$. 
Although little is known about properties of skewed parton distributions 
from first principles, at least the asymptotic 
distribution $G(u,\xi;\mu \to \infty)$ 
fulfills this  requirements \cite{Rad98b}.

The polarized gluon distribution enters
the production amplitude (\ref{eq:amplitude_deltaG}) 
via the integral
\begin{equation} 
\label{eq:delta_G_int}
\int_{-1}^{1}du\, \Delta G(u,\xi) 
\left[ \frac{1}{(\xi-u-i \epsilon)^2}-
\frac{1}{(\xi+u-i \epsilon)^2} \right] \, .
\end{equation}
Also this  integral exists only if $\Delta G(u,\xi)$ and its derivative are
continuous at $u = \xi$ which, again, is suggested by the asymptotic 
solutions of the corresponding  QCD evolution equations.

The dependence on the skewed gluon distribution 
in eq.(\ref{eq:delta_G_int}) 
is identical to the one found in $J/\Psi$ production
\cite{VantMank98}, 
but at variance with the model proposed in \cite{Rys97}. 
Integrating  eq.(\ref{eq:delta_G_int}) by parts shows that 
${\cal M}^{\gamma^*_\perp \to \rho_\perp}_{\Delta G}$
depends on the derivative of $\Delta G(u,\xi)$ rather then on $\Delta
G(u,\xi)$ itself. The imaginary part of 
${\cal M}^{\gamma^*_\perp \to \rho_\perp}_{\Delta G}$ 
is, for example, proportional to
\begin{equation}\label{eq:Im_part_amplitude_deltaG}
{\rm{Im}} \, \int_{-1}^{1}du\, \Delta G(u,\xi) 
\left[ \frac{1}{(\xi-u-i \epsilon)^2}-
\frac{1}{(\xi+u-i \epsilon)^2} \right] = - 2 \pi \,
\frac{\partial}{\partial u}\left.\Delta G(u,\xi)\right|_{u=\xi} \, .
\end{equation}
It can be argued that $\Delta G(u,\xi)$ is  
proportional to $u \Delta g(u)$ for small $\xi$ and $u \gsim \xi$ 
\cite{MPW98,Rad98c}. 
The magnitude of the imaginary part
(\ref{eq:Im_part_amplitude_deltaG}) in the small-$\xi$
region depends then crucially on the small-$u$ behavior of 
$\Delta g(u)$. 
Due to the derivative in eq.(\ref{eq:Im_part_amplitude_deltaG}) 
the contribution 
of polarized gluons to $\rho$ production is small if
$\Delta g(u)$ has a a strong singularity for small $u$  
as suggested e.g. in \cite{Bart96}. 
To the contrary, the contribution of polarized gluons can be large 
only if $\Delta g(u)$ does not rise fast at small $u$. 
This observation disagrees with the model calculation in ref.\cite{Rys97}.

So far we have considered only the gluonic part of the production
amplitude (\ref{eq:production_amplitude}). The quark amplitudes 
${\cal M}^{\gamma^*_\perp \to \rho_\perp}_q$ and 
${\cal M}^{\gamma^*_\perp \to \rho_\perp}_{\Delta q}$ 
have a form similar to the gluon ones,  but 
involve skewed unpolarized and polarized quark distributions 
$F(u,\xi)$ and $\Delta F(u,\xi)$. 
Both,  unpolarized and polarized quark 
distributions enter through integrals involving 
denominators $(\xi \pm u-i \epsilon)$ and their square 
as in (\ref{eq:G_old},\ref{eq:G_new},\ref{eq:delta_G_int}). 
However, for the asymptotic flavor singlet 
quark distribution \cite{Rad98b}
\begin{equation}\label{eq:q_asy}
F(u,\xi,\mu^2 \rightarrow \infty) = \frac{15}{2}\, \frac{N_F}{4 C_F + N_F} 
\frac{1}{\xi^2}\, \frac{u}{\xi}\left(1-\left(\frac{u}{\xi}\right)^2\right)\, 
\int_0^1 d\omega\,\left(\omega F_0(\omega,\xi) + G_0(\omega,\xi)
\right) 
\, ,
\end{equation}
one finds an important difference as compared to the 
gluon case: the real part of the integral  
\begin{equation} 
\label{eq:delta_F_int}
\int_{-1}^{1}du\, F(u,\xi) 
\left[ \frac{1}{(\xi-u-i \epsilon)^2}-
\frac{1}{(\xi+u-i \epsilon)^2} \right] \, ,
\end{equation}
which is present in ${\cal M}^{\gamma^*_\perp \to \rho_\perp}_q$, 
is infrared singular. 
It seems plausible that this  divergence occurs   
for any normalization scale $\mu$ \footnote{We thank A. Radyushkin 
for bringing this issue to our attention.}.
This observation makes clear that a phenomenological regularization of 
the integrals which involve  meson distribution amplitudes does not 
necessarily result in a finite result for the complete production 
amplitude.

In summary, we have shown via an explicit calculation 
of leading order Feynman diagrams how infrared singular 
contributions enter in the production of light vector mesons 
through the scattering of transversely polarized photons. 
They arise through integrals over involved meson 
production amplitudes and skewed parton distributions. 
The former  cannot be regularized through phenomenological meson distribution 
amplitudes with suppressed endpoint configurations. 
This is a direct consequence of Lorentz invariance 
which provides relations between different 
vector meson distributions.

In a phenomenological analysis of data on transverse 
$\rho$ meson production  model dependent regularizations  
have to be applied. 
If this procedure preserves 
the analytic structure suggested by a leading order calculation of  
Feynman diagrams, 
one obtains amplitudes which contain 
contributions from parton distribution functions  
and their derivatives. 
In particular polarized gluons enter  
only through their derivative.

\bigskip
 
\noindent
{\bf Acknowledgments:} 

\noindent
We would like to acknowledge discussions with 
V. Braun, A. Radyushkin, A. Sandacz, and W-D. Nowak.

\end{document}